\documentclass[preprint2]{aastex}

\slugcomment{Molecular Distributions in the Spiral Arm of M51}

\shorttitle{Molecular Distributions in the Spiral Arm of M51}
\shortauthors{Watanabe et al.}

\begin{document}


\title{Molecular Distribution in the Spiral Arm of M51}


\author{Yoshimasa~Watanabe\altaffilmark{1}, Nami~Sakai\altaffilmark{1,2}, Kazuo~Sorai\altaffilmark{3}, Junko Ueda\altaffilmark{4,5}}
\and
\author{Satoshi Yamamoto\altaffilmark{1}}
\email{nabe@taurus.phys.s.u-tokyo.ac.jp}


\altaffiltext{1}{Department of Physics, The University of Tokyo, 7-3-1 Hongo, Bunkyo-ku, Tokyo, 113-0033, Japan}
\altaffiltext{2}{RIKEN, 2-1, Hirosawa, Wako, Saitama 351-0198, Japan}
\altaffiltext{3}{Department of Physics / Department of Cosmoscience, Hokkaido University, Kita 10, Nishi 8, Kita-ku, Sapporo, Hokkaido, 060-0810, Japan}
\altaffiltext{4}{National Astronomical Observatory of Japan, 2-21-1 Osawa, Mitaka,Tokyo 181-8588, Japan}
\altaffiltext{5}{Harvard-Smithsonian Center for Astrophysics, 60 Garden Street, Cambridge, MA 02138, USA}


\begin{abstract}
Molecular line images of $^{13}$CO, C$^{18}$O, CN, CS, CH$_3$OH, and HNCO are obtained toward the spiral arm of M51 at a $ 7'' \times 6''$ resolution with the Combined Array for Research in Millimeter-wave Astronomy (CARMA).  Distributions of the molecules averaged over a 300~pc scale are found to be almost similar to one another and to essentially trace the spiral arm.  However, the principal component analysis shows a slight difference of distributions among molecular species particularly for CH$_3$OH and HNCO.  These two species do not correlate well with star-formation rate, implying that they are not enhanced by local star-formation activities but by galactic-scale phenomena such as spiral shocks.  Furthermore, the distribution of HNCO and CH$_3$OH are found to be slightly different, whose origin deserves further investigation.  The present results provide us with an important clue to understanding the 300~pc scale chemical composition in the spiral arm and its relation to galactic-scale dynamics.  
\end{abstract}





\section{Introduction}

Recently, various molecular species have readily been detected in nearby galaxies owing to rapid improvements in sensitivities of radioastronomical observations, and chemistry of molecular gas in external galaxies has attracted more attention of astronomers than before.   So far, about 60 molecular species have been identified in external galaxies.  Chemical studies for external galaxies have mostly been focused on nuclear regions, which usually give bright molecular emissions.  Spectral line surveys have extensively been conducted in such regions to characterize chemical nature of molecular gas \citep[e.g.][]{martin2006,aladro2011,aladro2013,Aladro2015,costagliola2011,nakajima2011,Snell2011}.  Distributions of various molecular species around the nuclear regions have been revealed with radio interferometers such as ALMA, and they are used to investigate peculiar physical states of the nuclear regions \citep[e.g.][]{Izumi2013,Sakamoto2014,Takano2014,Meier2015,Martin2015,Nakajima2015}.  

On the other hand, there are relatively few chemical studies in disk regions of external galaxies, because intensities of molecular line emissions are usually much weaker there than in the galactic nuclei.  \citet{Meier2005,Meier2012} imaged distributions of about 10 molecular species in the nuclear and bar regions of IC~342 and Maffei~2 with the OVRO and BIMA, and revealed significant chemical differentiation among giant molecular clouds (GMCs).  For instance, CH$_3$OH, HNCO, and SiO are enhanced in the bar region.  They concluded that large-scale shocks induced by gas dynamics specific to the bar structure are responsible to the enhancements.   However, chemical compositions observed toward external galaxies are those averaged over one or more GMCs within a telescope beam.  Therefore, we need to pay particular attention to this point, when we discuss chemical compositions of external galaxies on the basis of astrochemical concepts established for sub-pc scale regions in Galactic molecular clouds.  For instance, chemical evolutionary effects \citep[e.g.][]{Suzuki1992}, which are important for sub-pc scale clouds, may not be very important in the GMC-scale chemical compositions, because the time scale of the chemical equilibrium (10$^6$ yr) is much shorter than the sound crossing time of GMCs ($>10^7$ yr).  Hence, it seems likely that the GMC-scale chemical compositions are more or less in chemical equilibrium, and mainly depend on structures and environmental conditions of GMCs.  For better understanding of chemical effects in starbursts, AGNs, and shocks in spiral arms and bar regions, we need to establish the `standard' chemical composition of GMCs without or almost without these specific effects.  
With this motivation, we have been studying the chemical composition in the spiral arm region of M51 \citep{Watanabe2014}.

M51 is a grand-design spiral galaxy located at the distance of 8.4~Mpc \citep[][]{feldmeier1997,vink12}.   The distribution and dynamics of molecular gas in this galaxy have extensively been studied using both single dish telescopes \citep{nakai1994,schuster2007,Kuno2007,Miyamoto2014} and interferometers \citep{aalto1999,helfer2003,koda09,Schinnerer2013}.  Spectral line survey observations have been conducted toward two positions in the spiral arm in the 2~mm and 3~mm bands \citep{Watanabe2014} as well as the nuclear region in the 3~mm band \citep{Aladro2015}.  The spectral pattern observed in the two positions (P1 and P2) were very similar to each other indicating similar chemical compositions, although the one (P1) shows higher star formation activity than the other (P2).  Toward the P1 position, 15 molecular species have been identified.  The spectrum pattern is much different from that reported for high-mass star forming regions such as Orion KL \citep[e.g.][]{Tercero2010,Watanabe2015}, indicating that the observed chemical composition cannot simply be interpreted in terms of a composite of contributions from embedded star forming cores.  The GMC-scale distribution of molecules would mainly contribute to the spectrum.  Indeed, we detected the cold dense gas tracer N$_{2}$H$^{+}$ in our observation toward M51, which is relatively weak in high-mass star forming regions \citep{Watanabe2015}.  Assuming optically thin and local thermodynamic equilibrium (LTE) conditions, the excitation temperatures of CS, HNCO, and CH$_3$OH are estimated to be less than 10~K in M51~P1.  This result indicates that most of detected molecules reside in a cold ($\sim 10$~K) and widespread molecular gas, although a part of the molecular emission may also come from the hot molecular gas, \textit{e.g.} hot cores affected by the feedbacks from star formation activities.

However, the resolution of our previous line-survey observation is about $30''$, which corresponds to the linear scale of 1~kpc.  High angular resolution observations are needed to explore the origin and the nature of the molecular emission in more detail.  In this study, we imaged the 6 molecular species in the spiral arm of M51 with the Combined Array for Research in Millimeter-wave Astronomy (CARMA).  

\section{Observations with CARMA}
\begin{figure}
\epsscale{0.40}
\plotone{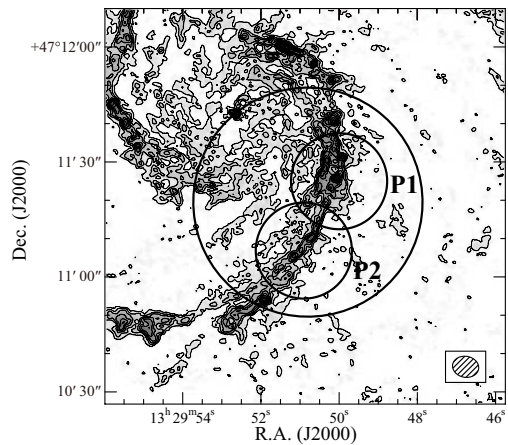}
\caption{A field of view of the observation with the CARMA (large circle) superposed on contours of the CO integrated intensity \citep[PAWS:][]{Schinnerer2013}.  Small circles indicate the beam size ($25''$) of the IRAM~30~m telescope at the 100~GHz toward the P1 and P2 positions \citep{Watanabe2014}.  A hatched ellipse indicates a synthesized beam ($6.9'' \times 5.5''$ with the position angle of $79.9^{\circ}$) of the CARMA observation at the 97~GHz.}
\label{fig_pos}
\end{figure}

The observations were carried out with the CARMA in May and June 2014.  It consists of six 10.4~m and nine 6.1~m antennas.  The primary beams of the 10.4~m and 6.1~m antennas at 110~GHz are about $60''$ and $100''$, respectively.  The phase-center coordinate is : (R.~A., Dec.) = (13:29:50.8, +47:11:19.5) in J2000 (Figure~\ref{fig_pos}).  Six molecular species, $^{13}$CO, C$^{18}$O, CN, CH$_3$OH, HNCO, and CS (Table \ref{tab_obs}), were simultaneously observed in the array configuration of D and E.  These configurations cover a baseline range of $\sim 2-55$~k$\lambda$.  The system noise temperatures were about 160--380~K.  The CH$_3$OH and CS lines and the C$^{18}$O, HNCO, $^{13}$CO, and CN lines were observed in the lower sideband and the upper sideband, respectively.  We employed 6 correlators for the 6 molecular lines.  The bandwidth of each correlator is 125~MHz with a frequency resolution of 0.781~MHz.  The bandpass calibration was done with the radio sources 3C273 and 3C279.  1419+543 was observed for 3~minutes every 15~minutes as a phase and gain calibrator.  The absolute flux of 1419+543 was measured by comparing with the flux of Mars.  The uncertainty of flux calibration is 20~\%.  The data reduction and analysis were done by using the MIRIAD package.  The synthesized beam size and the root-mean-square (r.~m.~s.) noise level of each molecule are summarized in Table~\ref{tab_obs}.  In the imaging procedure, the spectral channels were regrided.  The final velocity resolutions are summarized in Table~\ref{tab_obs}. 

\section{Results}
\subsection{Overview of molecules}
\begin{figure}
\epsscale{0.90}
\plotone{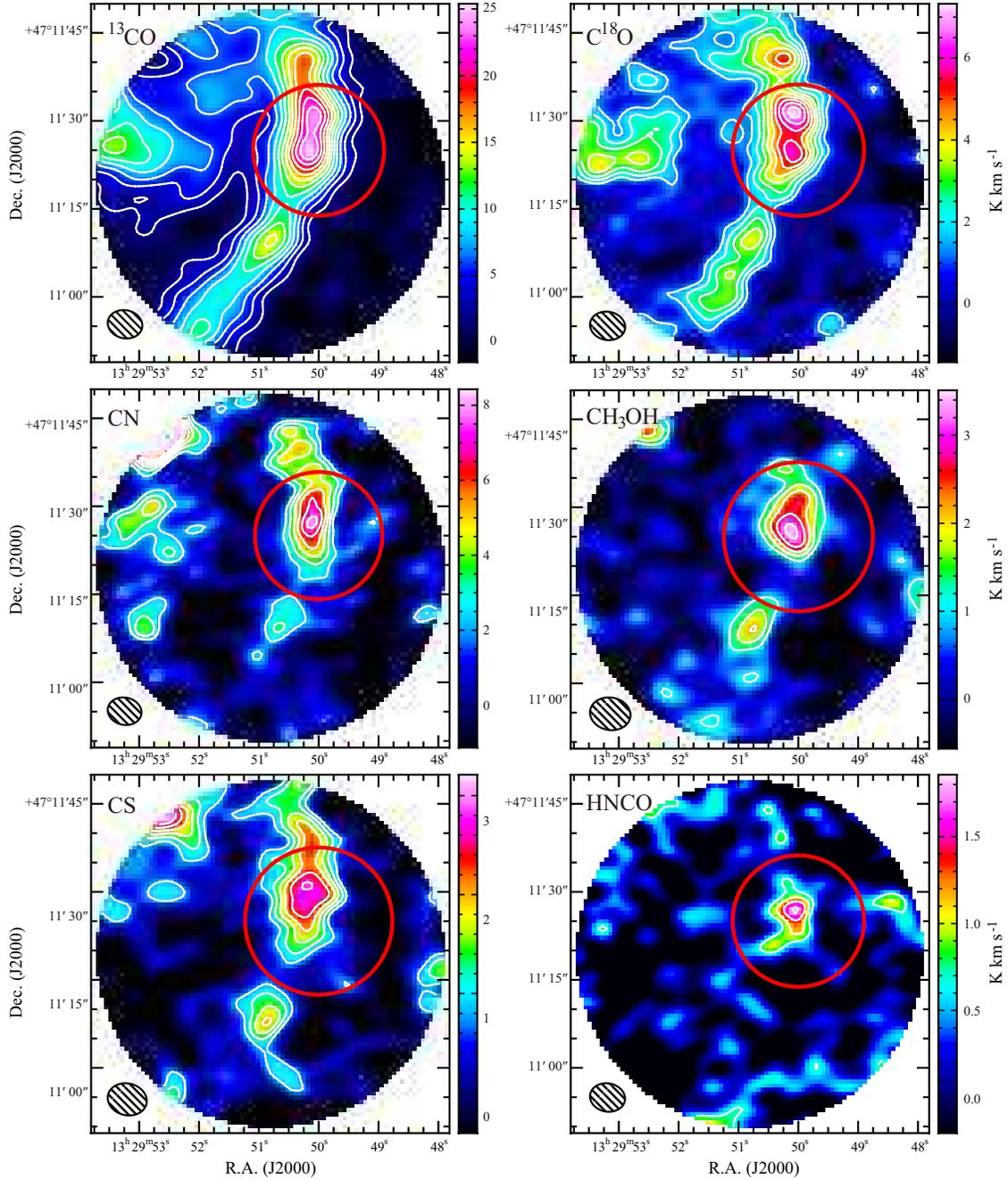}
\caption{Integrated intensity maps of $^{13}$CO, C$^{18}$O, CN, CH$_3$OH, CS, and HNCO.  The synthesized beams are shown as ellipticals in the lower left corner.  The red circles indicate the beam size of the IRAM 30~m telescope at P1.  Contour levels of the $^{13}$CO integrated intensity are 0.7~K~km~s$^{-1} \times (3, 6, 9, 12, 15, 18, 21, 24, 27, 30, 33)$.  Contour levels of the C$^{18}$O integrated intensity are 0.7~K~km~s$^{-1} \times (2, 3, 4, 5, 6, 7, 8, 9, 10)$.  Contour levels of the CN integrated intensity are 1.1~K~km~s$^{-1} \times (2, 3, 4, 5, 6, 7)$.  Contour levels of the CH$_3$OH integrated intensity are 0.55~K~km~s$^{-1} \times (2, 3, 4, 5, 6)$.  Contour levels of the CS integrated intensity are 0.40~K~km~s$^{-1} \times (2, 3, 4, 5, 6, 7)$.  Contour levels of the HNCO integrated intensity are 0.35~K~km~s$^{-1} \times (2, 3, 4, 5)$.  }
\label{fig_int}
\end{figure}

The $^{13}$CO ($J=1-0$), C$^{18}$O ($J=1-0$), CN ($N=1-0$), CH$_3$OH($J=2_{\rm k}-1_{\rm k}$), HNCO($5_{0\,5}-4_{0\,4}$), and CS($J=2-1$) lines were successfully detected.  Figure~\ref{fig_int} shows integrated intensity maps of the 6 molecular species.  A primary beam correction was applied to these maps.  The $^{13}$CO emission is found to be widely distributed along the spiral arm, nuclear bar, and molecular ring regions, which are defined by \citet{Hughes2013}.  Our $^{13}$CO($J=1-0$) map is similar to the $^{13}$CO($J=1-0$) map observed with the Owens Valley Radio Observatory mm-interferometer \citep{Schinnerer2010}, although the synthesized beam size of our map is coarser than their map (2.9$''$).  Since the intensity of the $^{13}$CO line is the strongest among the 6 molecules and the critical density is as low as $10^{3}$~cm$^{-3}$, this line would trace an entire distribution of molecular gas including relatively diffuse gas which cannot be traced by the other lines with higher critical densities.  The distribution of C$^{18}$O is more clumpy than that of $^{13}$CO.  The C$^{18}$O emission traces higher column-density regions, because the optical depth of the C$^{18}$O line is generally thinner than the $^{13}$CO line.  

Critical densities of the CN, CH$_3$OH, HNCO, and CS lines ($\sim 10^{5}$~cm$^{-3}$) are much higher than those of the $^{13}$CO and C$^{18}$O lines, and hence, these lines tend to trace denser molecular gas than the CO isotopologue lines.  The CN emission is detected along the spiral arm and shows a strong peak at the P1 position.  The CN distribution is extended from the CN peak to the northern direction.  The CS emission is also distributed along the spiral arm, and extends to the northern direction from the peak at the P1 position.  On the other hand, CH$_3$OH shows a rather compact distribution without any extended structures toward the northern direction from the peak at the P1.  HNCO shows a single peak at the P1 position probably due to an insufficient sensitivity of this observation.  The HNCO peak coincides with the CH$_3$OH peak.  Although the distributions of all the molecules are along the spiral arm, we recognize a slight variation among molecular species.

\subsection{Missing flux}
\begin{figure}
\epsscale{1.00}
\plotone{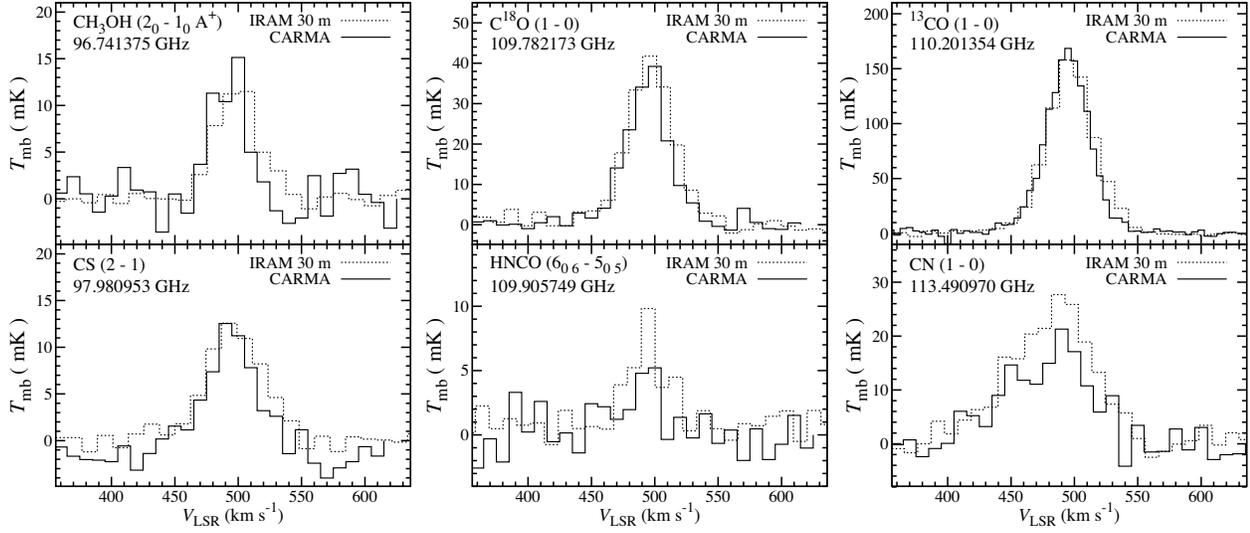}
\caption{Spectral line profiles observed toward the M51~P1 with the CARMA (solid line) and the IRAM~30~m telescope \citep[dotted line:][]{Watanabe2014}. The CARMA spectra are prepared by convolving the cleaned map by the resolution of the CH$_3$OH map. }
\label{fig_mf}
\end{figure}

Missing fluxes in the CARMA observations are evaluated by comparing with the fluxes obtained with the IRAM~30~m telescope \citep{Watanabe2014}.  The angular resolution of the CARMA maps are adjusted by convolving the Gaussian beam to match the beam size of the IRAM~30~m telescope.  The convolution is carried out by using a task \textit{convol} of the MIRIAD software package.  The angular resolution of the IRAM~30~m observation for each molecular line is summarized in Table~\ref{tab_intp1}.  Figure~\ref{fig_mf} shows the spectra of CH$_3$OH, CS, C$^{18}$O, HNCO, $^{13}$CO, and CN toward M51~P1 prepared by convolving the CARMA data and the corresponding spectra obtained with the IRAM~30~m telescope.  The spectral profile and intensity of the convolved spectrum is similar to the single dish spectrum for each molecule, although the spectral lines of HNCO and CN obtained with CARMA are significantly weaker than those with the IRAM~30~m telescope.  The integrated intensities of the both observations are shown in Table~\ref{tab_intp1}.  If uncertainties of intensity calibrations (20~\% both for the IRAM~30~m observations and the CARMA observations) are taken into account, most of fluxes are recovered in the CARMA observation, except for CN and HNCO.  As for HNCO, the missing flux is probably due to the poor S/N ratio in the CARMA observation.  On the other hand, the missing flux of CN, which is estimated to be 42~\%, is puzzling.  Since the missing flux of $^{13}$CO is small, this implies that CN is more extended than $^{13}$CO.  Weak CN emission extended outside the arm region may be resolved out in the CARMA observation.  Although we cannot in principle rule out the possibility that the missing flux of CN just originates from imperfect calibration, this possibility is unlikely, because the missing flux of most of molecules are negligible.  In any case, the CN abundances estimated in the latter sections tend to be lower than that estimated in \citet{Watanabe2014}.  

\subsection{Spectra at C$^{18}$O peaks}
\begin{figure}
\epsscale{0.40}
\plotone{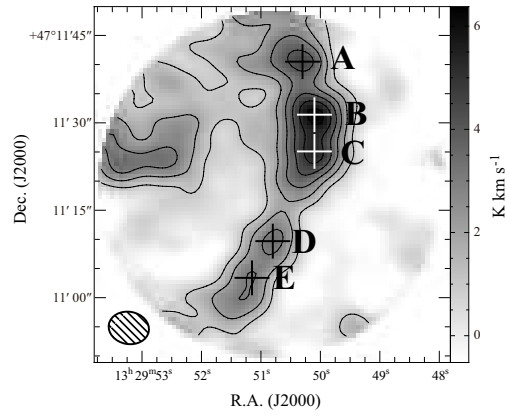}
\caption{Positions of the C$^{18}$O peaks (cross marks) superposed on the C$^{18}$O integrated intensity map.  Contour levels are 0.5~K~km~s$^{-1} \times (2, 4, 6, 8, 10, 12)$.  The angular resolution of the C$^{18}$O map is set to be the same as that of the CH$_3$OH map by convolution. }
\label{fig_lp}
\end{figure}

\begin{figure}
\epsscale{1.00}
\plotone{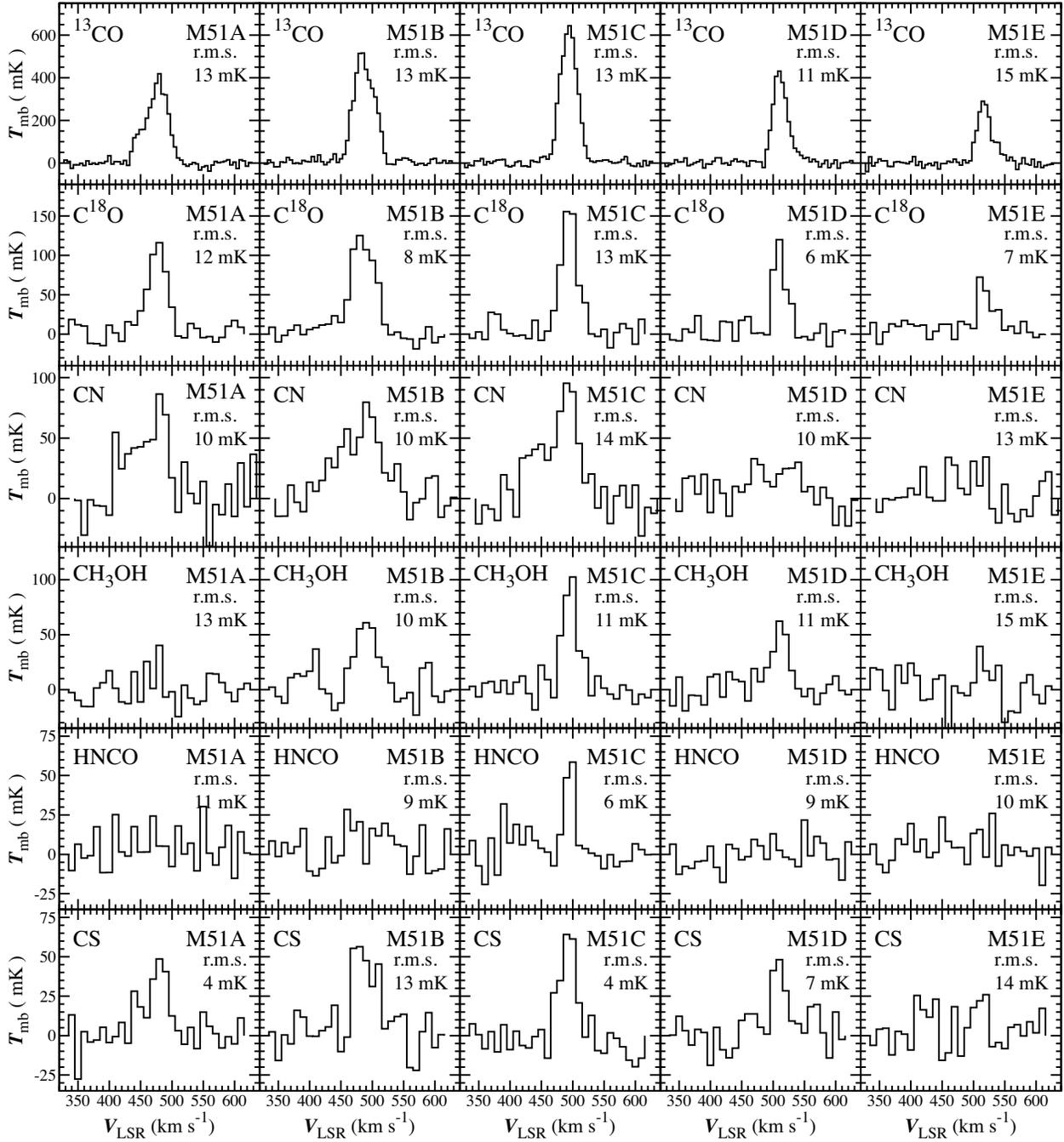}
\caption{Spectra of the 6 molecules sampled at the 5 C$^{18}$O peak positions (Table~\ref{tab_pos}). The spectra are obtained after the spatial resolution is converted to the resolution of the CH$_3$OH map ($6.0'' \times 6.9''$ with the position angle of $79.9^{\circ}$), which is the beam size of the CH$_3$OH map.  {\rm The angular resolution of CH$_3$OH map is used as the reference resolution, since the angular resolution is the coarsest.}  The intensity scale is the main beam temperature.  }
\label{fig_pspec}
\end{figure}

Spectral line parameters are derived for the five peaks of the C$^{18}$O integrated intensity (Figure~\ref{fig_lp} and Table~\ref{tab_pos}), because the C$^{18}$O peaks coincide with most of the other molecular line peaks and likely represent association of GMCs.  In order to compare the spectra at the same angular resolution, the $^{13}$CO, C$^{18}$O, CN, CS, and HNCO maps are convolved with the Gaussian beam for the CH$_3$OH map ($6.9'' \times 5.5''$) by using the MIRIAD task \textit{convol}.  The angular resolution of the final maps is thus the same as that of CH$_3$OH.  Figure~\ref{fig_pspec} shows the spectra obtained at the five C$^{18}$O peaks.  Table~\ref{tab_lineparm} summarizes peak intensities, integrated intensities, line-of-sight velocities, and line widths of the spectra.  
In the case of non-detection, the $3 \sigma$ upper limits to the peak temperature and the integrated intensity are given.  The line profiles of CN are different from those of the other molecules, because four hyperfine lines are blended.  Therefore, the linewidths and the line-of-sight velocities are not evaluated for the CN lines. 

\subsection{Molecular abundances}
Beam-averaged column densities of $^{13}$CO, C$^{18}$O, CN, CH$_3$OH, HNCO, and CS are evaluated for the C$^{18}$O peak positions (Table~\ref{tab_pos}), under the LTE assumption by using the following formula:
\begin{equation}
W_{\nu} = \frac{8 \pi^3 S\mu_0^2 \nu N}{3 k U(T_{\rm rot})} \left\{ 1 - \frac{\exp(h \nu/kT_{\rm rot})-1}{\exp(h \nu/kT_{\rm bg})-1} \right\}\exp\left(-\frac{E_{\rm u}}{k T_{\rm rot}}\right),
\end{equation}
where $W_{\nu}$, $S$, $\mu_0$, $\nu$, $N$, $k$, $T_{\rm rot}$, $U(T_{\rm rot})$, $h$, $T_{\rm bg}$, and $E_{\rm u}$ are integrated intensity, line strength, dipole moment, transition frequency, total column density, the Boltzmann constant, rotation temperature, partition function, the Planck constant, the cosmic microwave background temperature, and upper state energy, respectively.  The rotation temperature is assumed to be 10~K on the basis of the single dish result for a few molecules \citep{Watanabe2014}.  Since the beam sizes of all molecular data are the same as that of CH$_3$OH (see Section 3.3), a correction of a beam filling factor is not applied for each molecular line.  Hence, these column densities are ones averaged over the $6.0'' \times 6.9''$ scale.  The result is summarised in Table~\ref{tab_col}, and the errors are estimated from the rms noise of the spectrum and the calibration uncertainty (20~\%). The column densities are sensitive to the assumed rotation temperature.  When we assume the rotation temperature of 5~K and 20~K, the derived column densities vary by factor of 3 and 2, respectively, for the largest cases.  In the following analyses, we discuss the fractional abundances relative to H$_2$ instead of the column densities in order to mitigate the systematic errors caused by the assumption of the rotation temperature.  

Fractional abundances relative to H$_2$ are derived for the C$^{18}$O peak positions by using H$_2$ column densities estimated from C$^{18}$O, where the [H$_2$]/[C$^{18}$O] ratio of $2.9 \times 10^{6}$ is assumed \citep{Meier2005}.  The fractional abundances are summarised in Table~\ref{tab_fra}.  They are less affected by the assumed rotation temperature than the column densities.  Indeed, a difference of the fractional abundance is estimated to be at most within a factor of 2 for the rotation temperature range from 5~K to 20~K, where the maximum case is for HNCO.

\section{Discussion}
\subsection{Principal component analysis}
\begin{figure}
\epsscale{0.40}
\plotone{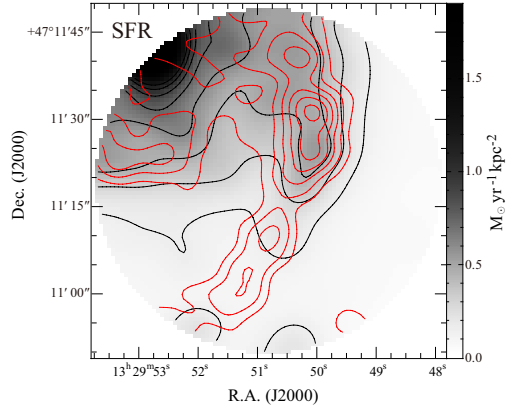}
\caption{A SFR map (gray scale and black contours) and an integrated intensity map of C$^{18}$O (red contours).  Contour levels are 0.15~M$_{\odot}$yr$^{-1}$kpc$^{-2} \times (1, 2, ..., 8)$ and 0.5~K~km~s$^{-1} \times (2, 4, ..., 12)$ for the SFR and the C$^{18}$O integrated intensity, respectively.  The angular resolutions of the two maps are convolved to the resolution of the CH$_3$OH map ($6.0'' \times 6.9''$ with the position angle of $79.9^{\circ}$). }
\label{fig_sfrmap}
\end{figure}

We conduct a principal-component analysis (PCA) to evaluate similarity and difference among maps of different molecular lines quantitatively.  The PCA has been used as a common analytical technique to quantify morphological correlation of molecular distributions in molecular clouds and external galaxies \citep[e.g.][]{Ungerechts1997,Meier2005}.  In addition to the molecular distributions obtained with the CARMA, a surface density distribution of the star-formation rate (SFR) is included in the PCA.  The SFR is estimated from the intensities of the H$\alpha$ and 24~$\mu$m emission observed by \citet{Kennicutt2003}.  In order to compare the SFR map with our molecular maps, the angular resolutions of the H$\alpha$ and 24~$\mu$m maps are convolved to be the resolution of the CH$_3$OH map.  After the convolution, the SFR is evaluated by using the method given by \citet{Calzetti2007}, as shown in Figure~\ref{fig_sfrmap}.  We used ${\rm SFR}(M_{\odot}\, {\rm yr}^{-1}) = 5.3 \times 10^{-42} [L({\rm H \alpha})_{\rm obs} + (0.031 \pm 0.006) \times L({\rm 24 \, \mu m})] $, where $L({\rm H \alpha})_{\rm obs}$ and $L({\rm 24 \, \mu m})$ are observed luminosities of H$\alpha$ and 24~$\mu$m in erg~s$^{-1}$, respectively. 

\begin{figure}
\epsscale{0.90}
\plotone{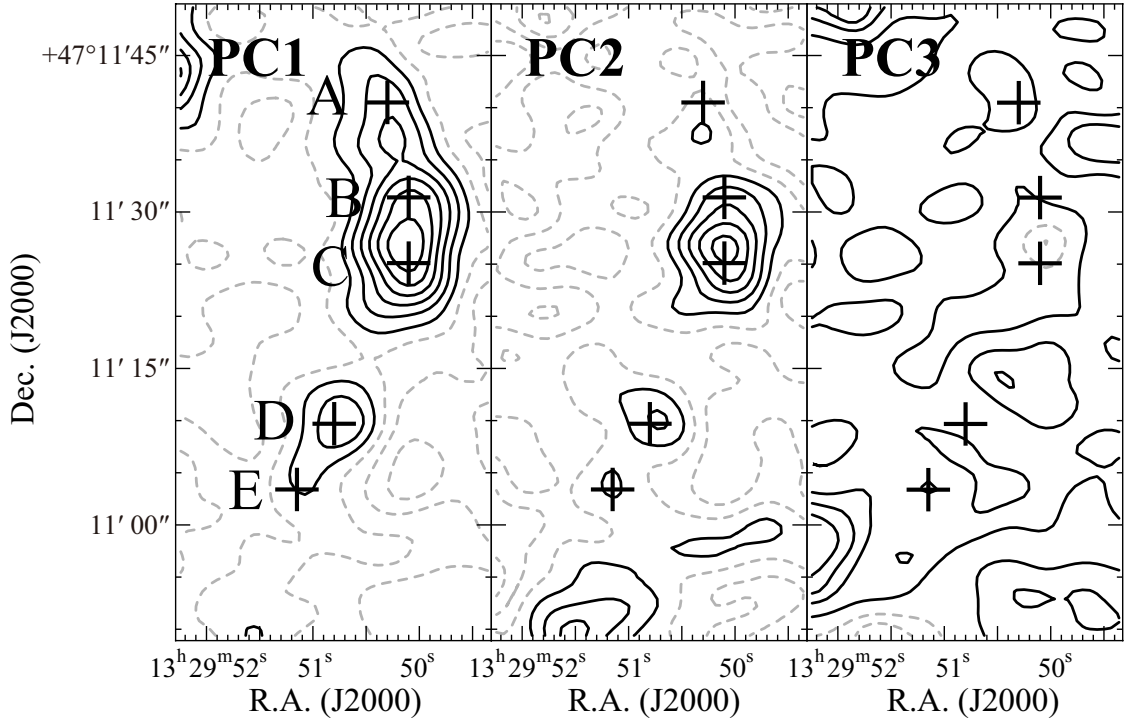}
\caption{Maps of the first, second, and third principal components.  The black contours and dashed contours indicate positive and negative values, respectively.}
\label{fig_pca_map}
\end{figure}

The PCA is conducted in the same way as \citet{Meier2005} reported.  The molecular line maps are convolved and re-gridded to the same angular resolution for the CH$_3$OH emission.  Then, the pixel values of the molecular maps and the SFR are normalized and mean-centered (Appendix~A).  After these preprocesses, a correlation matrix is calculated from the preprocessed maps, and its eigenvalues and eigenvectors are then calculated by diagonalizing the correlation matrix.  Tables~\ref{tab_corr} and \ref{tab_eigv} show the correlation matrix and the eigenvectors, respectively.  Figure~\ref{fig_pca_map} shows the maps of the first, second and third principal components. 

\begin{figure}
\epsscale{0.90}
\plotone{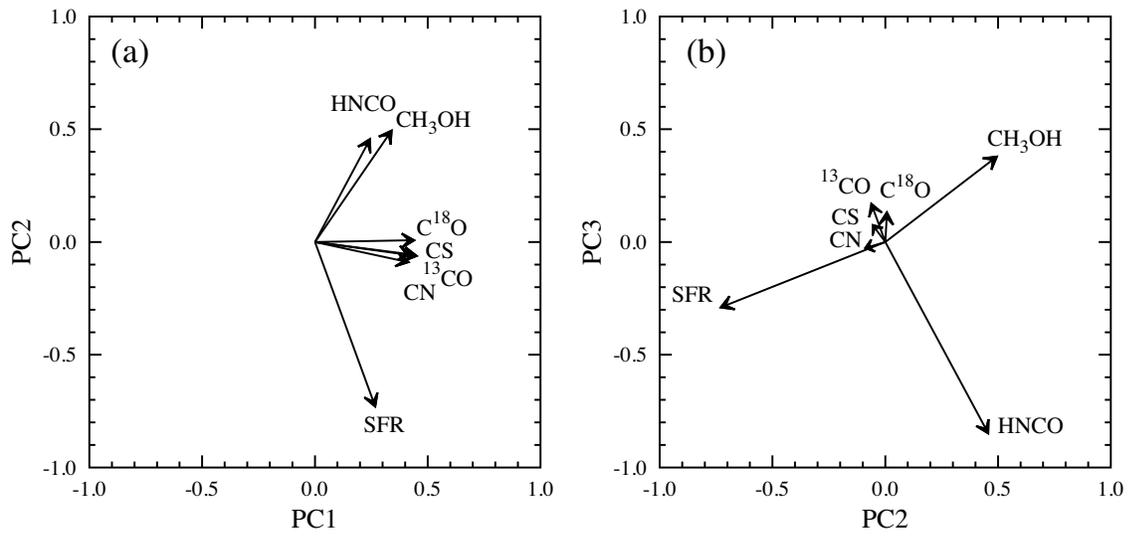}
\caption{The projection of each molecule on (a) PC1--PC2 and (b) PC2--PC3 planes.  The meanings of PC1, PC2, and PC3 are explained in Section 4.1.}
\label{fig_pca}
\end{figure}

Figure~\ref{fig_pca} shows projection of each species on the first, second, and third PC axes.  On the first principal component (PC1) axis, all the species have positive projections with a similar magnitude (0.24--0.45).  Hence, the PC1 represents an averaged distribution of molecules and SFR, which essentially traces the spiral arm structure (Figure~\ref{fig_pca_map}).  On the second and third principal component (PC2 and PC3, respectively) axes, $^{13}$CO, C$^{18}$O, CS, and CN have relatively smaller projection than the other molecules.  The PC2 mainly highlights the CH$_3$OH, HNCO, and SFR distributions, which is characteristic in distributions around the peak C of C$^{18}$O (PC2 in Figure~\ref{fig_pca_map}).  The SFR has an opposite sign to those of CH$_3$OH and HNCO on the PC2 axis, indicating that the distributions of the SFR and the CH$_3$OH--HNCO group tend to be anticorrelated.  The PC3 characterizes the different distributions between CH$_3$OH and HNCO.   The PC2-PC3 plane shows that CN is the only molecule which has same direction of SFR, although the magnitude of CN is relatively small.  In this way, the PCA sensitively shows the slight difference among the distributions of the molecules and the SFR at a 300~pc scale in the spiral arm.  

\subsection{Radial distribution of abundance}
\begin{figure}
\epsscale{0.40}
\plotone{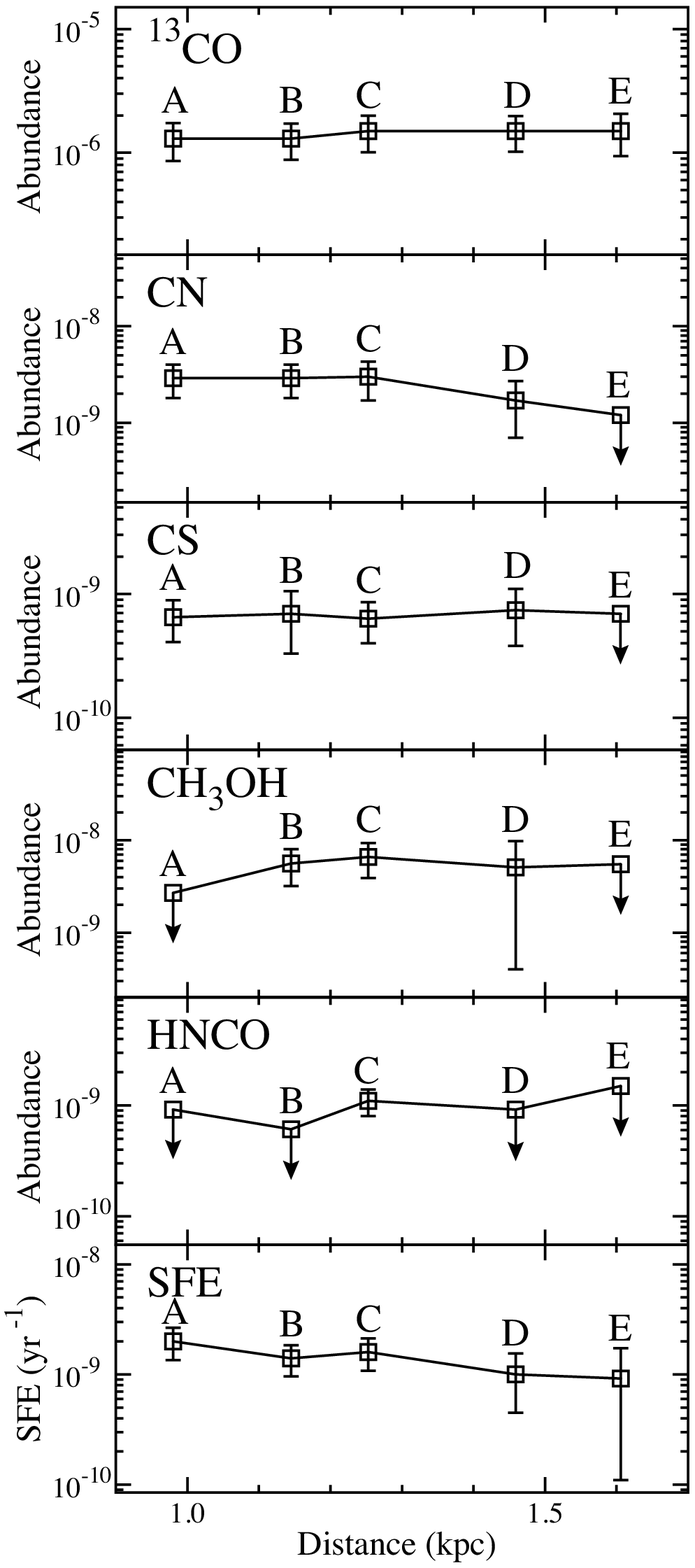}
\caption{Fractional abundances of CN, CS, CH$_3$OH, and HNCO relative to H$_2$ as a function of the galactocentric distance (top five figures).  The star formation efficiency (SFE) as a function of the garactocentric distance (bottom figure).  Squares with arrow indicate upper limits.}
\label{fig_XSFEdist}
\end{figure}

Figure~\ref{fig_XSFEdist} shows the fractional abundances of $^{13}$CO, CN, CS, CH$_3$OH, and HNCO relative to H$_2$, and the star formation efficiencies (SFE) which are derived by dividing the SFR by the H$_2$ gas mass, at the C$^{18}$O peak positions. They are indicated as a function of the galactocentric distance, where the galactocentric distances are estimated by assuming the distance to M51 of 8.4~Mpc, an inclination angle of 22$^{\circ}$, and a position angle of 173$^{\circ}$ \citep{Colombo2014}.  The fractional abundances of $^{13}$CO and CS are found to be almost constant over the observed regions along the spiral arm.  On the other hand, the fractional abundances of the other molecules and the SFE slightly depends on positions.  The fractional abundances of CN and the SFE are higher at Positions A, B, and C than at D and E by factor of 2--1.5.  The fractional abundance of CN and the SFE may be correlated with each other along the spiral arm, as the PCA shows a hint of correlation between the integrated intensity map of CN and the SFR distribution.  In contrast, the fractional abundance of CH$_3$OH is lower at Position A than at the other positions.  It apparently differs from the trend in the SFE.  

\subsection{CH$_3$OH}
In this study, we find that CH$_3$OH has a relatively high fractional abundance ($\sim 6 \times 10^{-9}$).  In addition, we also recognize that CH$_3$OH is distributed differently from the other molecules.  CH$_3$OH is thought to be formed on the cold dust mantle through hydrogenation of CO, since the CH$_3$OH formation in the gas phase is not efficient \citep{Garrod2007}.  Some liberation mechanisms are necessary for CH$_3$OH to be detected in the gas phase.  In the Galactic objects, the enhancement of CH$_3$OH is usually found in hot cores and hot corinos in star-forming regions \citep[e.g.][]{Jorgensen2005,Bisschop2007,Bachiller1997}.  Radiative heatings or/and shocks induced by star formation activities such as protostellar outflows are thought to be main desorption mechanisms of CH$_3$OH.  However, the observed CH$_3$OH abundance is not clearly correlated but slightly anticorrelated with the SFE in the spiral arm of M51.  This result is consistent with the result of our previous observation with the IRAM~30~m telescope at a 1~kpc scale \citep{Watanabe2014}.  No relevance of the CH$_3$OH enhancement to star formation activities at a GMC-scale was found in the bar regions of IC~342 and Maffei~2 \citep{Meier2005,Meier2012}, either.  Hence, it is most likely that the star formation feedback does not contribute to the gas-phase CH$_3$OH abundance in M51 at a 300~pc scale. 

\citet{Meier2005,Meier2012} concluded that the CH$_3$OH distribution traces large-scale shocks induced by gas orbital resonance in the bar regions.  Indeed, other shock tracers such as SiO are found to be enhanced in the bar regions \citep{Usero2006,Meier2012}.  Although our observed position of M51 is not the bar region, spiral shocks may occur in the spiral arm \citep[e.g.][]{Fujimoto1968,Roberts1969,Shu1972}.  Such a spiral shock could be responsible for evaporation of CH$_3$OH.  If the spiral shock causes the enhancement of CH$_3$OH, the non-uniform distribution of the CH$_3$OH would be originated from variation of shock effects in the spiral arm (i.e.  variation of shock strength and/or internal structure of the arm).  Another possible mechanism of the CH$_3$OH liberation is non-thermal desorption such as cosmic-ray induced UV photon, as discussed in \citet{Watanabe2014}.  The abundance of CH$_3$OH estimated in this observation is on the same order of that in the cold quiescent core TMC-1 ($\sim 10^{-9}$), where the non-thermal desorption is thought to be responsible for the widely distributed CH$_3$OH \citep{Soma2015}.  Hence, the non-thermal desorption could also explain the observed CH$_3$OH abundance in the spiral arm of M51.  

In addition to the liberation mechanisms, the efficiency of the CH$_3$OH formation on grain mantle would also affect the CH$_3$OH distribution in the spiral arm.  For dust temperature higher than 30~K, the CH$_3$OH formation is inefficient on grain mantle, because CO depletion does not occur above 20~K \citep[e.g.][]{Aikawa2008}.  \citet{Schinnerer2010} found that the gas kinetic temperature decreases with the galactocentric distance in M51 on the basis of the LVG analysis of their CO observation.  In our observation, the CH$_3$OH abundance is relatively low in Position A.  This position is the closest position to the nuclear region of M51, where the temperature may be higher than the other positions.  

Above all, the characteristic distribution of CH$_3$OH would be originated from a combination of evaporation mechanisms and formation mechanisms of CH$_3$OH.  For further understanding of CH$_3$OH in the spiral arm, sensitive multi-line observations of CH$_3$OH in other regions including inter-arm regions where the spiral shock does not occur, as well as detailed analyses of kinematics and physical conditions of molecular gas, are necessary.  These are left for future studies.

\subsection{Other molecules}
In this study, we find a hint that the fractional abundance of CN is correlated with the SFE.  If so, the CN production may be related to photodissociation of HCN by UV photons from star formation activities or from the nuclear region \citep[e.g.][]{Ginard2015}.  In this case, enhancement of other molecules related to the PDR (photodissociation region), such as CCH and CO$^+$, could be expected \citep[e.g.][]{Ginard2015,Fuente2006,Pety2005}.  Sensitive observations of these molecules are thus interesting.   

\citet{Meier2005,Meier2012} suggested that HNCO comes from grain mantle by the shock evaporation, because the distribution of HNCO is similar to that of CH$_3$OH in IC~342 and Maffei~2.  The fractional abundance of HNCO in M51 is similar to that in IC~342 ($(1-3) \times 10^{-9}$) \citep{Meier2005}.  The peak position of HNCO almost coincides with that of CH$_3$OH in M51, and hence, HNCO in our observation may also come from grain mantle.  On the other hand, our PCA suggests that the distribution of HNCO is slightly different from those of the other molecules including CH$_3$OH, although sensitivity of the HNCO observation is not good.  Indeed, the HNCO is detected only in one of five positions (C) although the signal-to-noise ratio is 3.2.  From observations of the Galactic GMCs, HNCO has been detected with relatively strong intensities in the GMCs near the Galactic center \citep[e.g.][]{Jackson1984,Armstrong1985,Cummins1986}.  The fractional abundance toward the hot core of Sgr~B2(N) is $1.6\times 10^{-7}$ \citep{Marcelino2010}, while that toward the hot core of Orion~KL is much lower ($9 \times 10^{-9}$) \citep{Zinchenko2000} than Sgr~B2(N).  The HNCO abundance is also known to be enhanced in the shocked region L1157~B1 ($(0.4-1.8) \times 10^{-8}$), where HNCO is liberated from grain mantle \citep{Rodoriguez2010}.  On the other hand, the HNCO abundance is as low as $(1-5) \times 10^{-10}$ in cold clouds \citep{Marcelino2009}, where the gas phase production is considered to be dominant.  The fractional abundances in M51 is much lower than abundances in Sgr~B2(N) and L1157~B1, while it is slightly higher than that in cold clouds.  Hence, evaporation of grain mantle seems to contribute to the gas phase abundance of HNCO.  Considering poor correlation with the SFR, a galactic shock may play an important role to some extent as in the case of CH$_3$OH.  Since the surface binding energy of HNCO (2850~K) is lower than that of CH$_3$OH (4930~K) \citep{McElroy2013}, the difference of their distribution may be due to shock strength.  It is suggested that HNCO is thought to be unstable under strong UV field \citep{Roberge1991,Martin2008}, and this may cause the slight difference between CH$_3$OH and HNCO distribution.  Thus, the meaning of the distribution of HNCO is left for future studies.  

\section{Concluding Remarks}
Our interferometric observations of the 6 molecular species toward the spiral arm of M51 at a spatial resolution of 300~pc reveal that the molecular distributions almost look similar to one another and mainly trace the spiral arm structure.  A detailed look at the distributions by the PCA and evaluation of the fractional abundances as a function of the galactocentric distance shows a slight chemical differentiation.  It should be noted that the CH$_3$OH distribution is not well correlated to the SFR.  Hence, the effect of the star formation and its feedback is not significant in the CH$_3$OH distribution averaged over the 300~pc scale under a mild star formation activity environment.  Similar results are obtained for HNCO and CS.  Rather the galactic scale phenomena occurring in the spiral arm, such as the spiral shocks, would be responsible for the slight chemical differentiation.  


\acknowledgments
The authors thank Susanne Aalto for invaluable discussions.  The authors are grateful to the CARMA staff for their excellent support. Support for CARMA construction was derived from the Gordon and Betty Moore Foundation, the Kenneth T. and Eileen L. Norris Foundation, the James S. McDonnell Foundation, the Associates of the California Institute of Technology, the University of Chicago, the states of California, Illinois, and Maryland, and the National Science Foundation.  CARMA development and operations are supported by the National Science Foundation under a cooperative agreement, and by the CARMA partner universities.  This study is supported by a Grant-in-Aid from the Ministry of Education, Culture, Sports, Science, and Technology of Japan (No. 21224002, 21740132, and 25108005).

\appendix
\section{Preprocessing Procedure in the PCA}
In order to compare different physical quantities of the integrated intensities of molecules and the star formation rate, the pixel values of a map are normalized and mean-centered as follows:
\begin{equation}
\widetilde{I}_{i\,j} = \frac{I_{i\,j} - \overline{I}_{i}}{\sigma_{i}},
\end{equation}
where $\widetilde{I}_{i\,j}$, $i$, $j$, $I_{i\,j}$, $\overline{I}_{i}$, and $\sigma_{i}$, are a processed pixel value, an index of molecules or star formation rate, an index of the pixel, a pixel value of the original map, a mean pixel value, and an unbiased estimate of variance of the pixel value, respectively.
The mean pixel value is given as:
\begin{equation}
\overline{I}_{i} = \frac{1}{n}\sum^{n}_{j=1} I_{i\,j}.
\end{equation}
where $n$ is the number of pixel.  The unbiased estimate of variance of the pixel value is denoted as:
\begin{equation}
\sigma^{2}_{i} = \frac{1}{n-1}\sum^{n}_{j=1} (I_{i\,j} - \overline{I}_{i})^2.
\end{equation}

\clearpage



\clearpage







\clearpage
\begin{deluxetable}{llrcccc}
\tabletypesize{\scriptsize}
\tablecaption{Summary of observation}
\tablewidth{0pt}
\tablehead{
\colhead{Molecule} & \colhead{Transition} & \colhead{Frequency} & \colhead{Sideband} & \colhead{Beam} \tablenotemark{a} & \colhead{Vel. Res.} \tablenotemark{b} & \colhead{r.m.s.} \tablenotemark{c} \\
\colhead{}         & \colhead{} & \colhead{(GHz)}     & \colhead{} & \colhead{}   & \colhead{(km~s$^{-1}$)} & \colhead{(mK~beam$^{-1}$)} 
}
\startdata
CH$_3$OH  & $2_0-1_0$ A$^+$       &  96.741375 & LSB & $6.9''\times5.5'', 79.9^{\circ}$ & 10.0 & 13.9 \\
CS        & $J=2-1$               &  97.980953 & LSB & $6.7''\times5.4'', 76.4^{\circ}$ & 10.0 & 14.7 \\
C$^{18}$O & $J=1-0$               & 109.782173 & USB & $6.0''\times4.8'', 79.6^{\circ}$ & 10.0 & 15.7 \\
HNCO      & $6_{0\,6} - 5_{0\,5}$ & 109.905749 & USB & $6.0''\times4.8'', 79.3^{\circ}$ & 10.0 & 16.0 \\
$^{13}$CO & $J=1-0$               & 110.201354 & USB & $6.0''\times4.8'', 77.6^{\circ}$ &  5.0 & 21.9 \\
CN        & $N=1-0\,J=3/2-1/2$    & 113.490970 & USB & $5.8''\times4.7'', 78.1^{\circ}$ & 10.0 & 24.9 \\
\enddata
\tablenotetext{a}{FWHM size of the major axis, FWHM size of the minor axis, and position angle of the synthesized beam.}
\tablenotetext{b}{Velocity resolution of spectrometer channels.}
\tablenotetext{c}{Sensitivity of the image at the velocity resolution listed in this table.}
\label{tab_obs}
\end{deluxetable}

\begin{deluxetable}{lrrrr}
\tabletypesize{\scriptsize}
\tablecaption{Integrated intensities in the M51~P1}
\tablewidth{0pt}
\tablehead{
\colhead{Molecule} & \colhead{Frequency} & \colhead{Resolution} \tablenotemark{a} & \colhead{CARMA} \tablenotemark{b} & \colhead{IRAM~30~m} \tablenotemark{b} \\
\colhead{}         & \colhead{(GHz)}     & \colhead{(arcsec)}   & \colhead{(K~km~s$^{-1}$)} & \colhead{(K~km~s$^{-1}$)} 
}
\startdata
CH$_3$OH  &  96.741375 & 25.4 & $0.47 \pm 0.14$ & $0.52 \pm 0.05$ \\
CS        &  97.980953 & 25.1 & $0.49 \pm 0.17$ & $0.68 \pm 0.08$ \\
C$^{18}$O & 109.782173 & 22.4 & $1.53 \pm 0.09$ & $1.89 \pm 0.14$ \\
HNCO      & 109.905749 & 22.4 & $0.13 \pm 0.11$ & $0.32 \pm 0.10$ \\
$^{13}$CO & 110.201354 & 22.3 & $7.06 \pm 0.19$ & $7.48 \pm 0.16$ \\
CN        & 113.490970 & 21.7 & $1.21 \pm 0.23$ & $2.08 \pm 0.05$ \\
\enddata
\tablenotetext{a}{The resolution of the IRAM 30~m telescope is calculated by $2460/\nu$ (arcsec), where $\nu$ is the observing frequency in GHz.}
\tablenotetext{b}{The errors are 3$\sigma$.}
\label{tab_intp1}
\end{deluxetable}

\begin{deluxetable}{lcc}
\tablecaption{Positions of the C$^{18}$O peaks \tablenotemark{a}}
\tablewidth{0pt}
\tablehead{
\colhead{Position} & \colhead{R.~A. (J2000)} & \colhead{Dec. (J2000)} 
}
\startdata
A       & 13$^{\rm h}$ 29$^{\rm m}$ 50$^{\rm s}$.3 & +47$^{\circ}$ 11$'$ 40$''$.5 \\
B       & 13$^{\rm h}$ 29$^{\rm m}$ 50$^{\rm s}$.1 & +47$^{\circ}$ 11$'$ 31$''$.4 \\
C       & 13$^{\rm h}$ 29$^{\rm m}$ 50$^{\rm s}$.1 & +47$^{\circ}$ 11$'$ 25$''$.1 \\
D       & 13$^{\rm h}$ 29$^{\rm m}$ 50$^{\rm s}$.8 & +47$^{\circ}$ 11$'$ ~9$''$.7 \\
E       & 13$^{\rm h}$ 29$^{\rm m}$ 51$^{\rm s}$.1 & +47$^{\circ}$ 11$'$ ~3$''$.4 \\
\enddata
\tablenotetext{a}{Positions (A--E) shown in Figure~\ref{fig_lp}.}
\label{tab_pos}
\end{deluxetable}

\begin{deluxetable}{llcccc}
\tabletypesize{\scriptsize}
\tablecaption{Line parameters at the C$^{18}$O peaks}
\tablewidth{0pt}
\tablehead{
\colhead{Position} &\colhead{Molecule} & \colhead{$T$ Peak} \tablenotemark{a} & \colhead{$\int T dv$} \tablenotemark{a} & $V_{\rm LSR}$ & FWHM \\
\colhead{} &\colhead{}         & \colhead{(mK)} & \colhead{(K~km~s$^{-1}$)} & \colhead{(km~s$^{-1}$)} & \colhead{(km~s$^{-1}$)}
}
\startdata
A      & $^{13}$CO & $419 \pm 38$ & $16.52 \pm 0.79$ & $480 \pm 5$  & $40 \pm 5$   \\
       & C$^{18}$O & $116 \pm 35$ &  $4.29 \pm 0.92$ & $480 \pm 10$ & $40 \pm 10$  \\
       & CN        &  $86 \pm 31$ &  $4.15 \pm 0.92$ &              &              \\
       & CH$_3$OH  &  $< 38$      &  $< 1.0$         &              &              \\
       & HNCO      &  $< 32$      &  $< 0.84$        &              &              \\
       & CS        &  $49 \pm 12$ &  $1.70 \pm 0.31$ & $480 \pm 10$ & $30 \pm 10$  \\
B      & $^{13}$CO & $516 \pm 38$ & $21.10 \pm 0.94$ & $485 \pm 5$  & $40 \pm 5$   \\
       & C$^{18}$O & $125 \pm 23$ &  $5.69 \pm 0.65$ & $480 \pm 10$ & $50 \pm 10$  \\
       & CN        &  $80 \pm 29$ &  $5.44 \pm 1.16$ &              &              \\
       & CH$_3$OH  &  $61 \pm 30$ &  $2.72 \pm 0.80$ & $490 \pm 10$ & $30 \pm 10$  \\
       & HNCO      &  $< 26$      &  $< 0.74$        &              &              \\
       & CS        &  $56 \pm 39$ &  $2.49 \pm 1.04$ & $480 \pm 10$ & $50 \pm 10$  \\
C      & $^{13}$CO & $644 \pm 38$ & $21.93 \pm 0.93$ & $495 \pm 5$  & $35 \pm 5$   \\
       & C$^{18}$O & $156 \pm 38$ &  $5.31 \pm 1.07$ & $490 \pm 10$ & $30 \pm 10$  \\
       & CN        &  $95 \pm 43$ &  $5.43 \pm 1.43$ &              &              \\
       & CH$_3$OH  & $103 \pm 31$ &  $3.02 \pm 0.71$ & $500 \pm 10$ & $20 \pm 10$  \\
       & HNCO      &  $58 \pm 16$ &  $1.29 \pm 0.40$ & $500 \pm 10$ & $20 \pm 10$  \\
       & CS        &  $64 \pm 13$ &  $2.08 \pm 0.30$ & $490 \pm 10$ & $30 \pm 10$  \\
D      & $^{13}$CO & $431 \pm 34$ & $12.70 \pm 0.64$ & $510 \pm 5$  & $25 \pm 5$   \\
       & C$^{18}$O & $120 \pm 18$ &  $2.97 \pm 0.36$ & $510 \pm 10$ & $20 \pm 10$  \\
       & CN        &  $33 \pm 31$ &  $1.72 \pm 0.89$ &              &              \\
       & CH$_3$OH  &  $62 \pm 33$ &  $1.65 \pm 0.74$ & $510 \pm 10$ & $30 \pm 10$  \\
       & HNCO      &  $< 26$      &  $< 0.58$        &              &              \\
       & CS        &  $48 \pm 22$ &  $1.35 \pm 0.49$ & $510 \pm 10$ & $30 \pm 10$  \\
E      & $^{13}$CO & $290 \pm 46$ &  $8.24 \pm 0.76$ & $515 \pm 5$  & $25 \pm 5$   \\
       & C$^{18}$O &  $72 \pm 22$ &  $1.87 \pm 0.44$ & $515 \pm 10$ & $20 \pm 10$  \\
       & CN        &  $<39$       &  $< 0.78$        &              &              \\
       & CH$_3$OH  &  $< 46$      &  $< 0.92$        &              &              \\
       & HNCO      &  $< 30$      &  $< 0.60$        &              &              \\
       & CS        &  $< 41$      &  $< 0.83$        &              &              \\
\enddata
\tablenotetext{a}{The errors are 3$\sigma$.}
\label{tab_lineparm}
\end{deluxetable}

\begin{deluxetable}{lcccccc}
\tabletypesize{\scriptsize}
\tablecaption{Column densities at the C$^{18}$O peaks}
\tablewidth{0pt}
\tablehead{
\colhead{Position} &\colhead{$^{13}$CO}   & \colhead{C$^{18}$O}   & \colhead{CN}          & \colhead{CH$_3$OH}    & \colhead{HNCO}        & \colhead{CS} \\
\colhead{}         &\colhead{$\times 10^{16}$ (cm$^{-2}$)} & \colhead{$\times 10^{15}$ (cm$^{-2}$)} & \colhead{$\times 10^{13}$ (cm$^{-2}$)} & \colhead{$\times 10^{13}$ (cm$^{-2}$)} & \colhead{$\times 10^{13}$ (cm$^{-2}$)} & \colhead{$\times 10^{13}$ (cm$^{-2}$)}
}
\startdata
A      & $1.6\pm0.3$ & $4.3\pm1.2$ & $3.5\pm1.0$ & $<3.3$      & $<1.1$      & $(0.8\pm0.2)$\\
B      & $2.1\pm0.4$ & $5.6\pm1.3$ & $4.6\pm1.3$ & $8.9\pm3.2$ & $<1.0$      & $(1.1\pm0.5)$\\
C      & $2.2\pm0.4$ & $5.3\pm1.5$ & $4.5\pm1.5$ & $9.9\pm3.1$ & $1.7\pm0.6$ & $(1.0\pm0.2)$\\
D      & $1.3\pm0.3$ & $2.9\pm0.7$ & $1.4\pm0.8$ & $5.4\pm2.7$ & $<0.8$      & $(0.6\pm0.3)$\\
E      & $0.8\pm0.2$ & $1.9\pm0.6$ & $< 0.7$     & $<3.0$      & $<0.8$      & $<0.4$\\
\enddata
\tablecomments{Errors of the column densities are estimated by taking into account the r.~m.~s. noise and calibration uncertainty (20~\%).  The column densities are calculated under the LTE approximation with a rotation temperature of 10 K.}
\label{tab_col}
\end{deluxetable}

\begin{deluxetable}{lccccc}
\tabletypesize{\scriptsize}
\tablecaption{Fractional abundances relative to H$_2$ at the C$^{18}$O peaks}
\tablewidth{0pt}
\tablehead{
\colhead{Position} & \colhead{$^{13}$CO} & \colhead{CN} & \colhead{CH$_3$OH} & \colhead{HNCO} & \colhead{CS} 
}
\startdata
A & $(1.3\pm0.4) \times 10^{-6}$ & $(2.9\pm1.1) \times 10^{-9}$ & $<2.7 \times 10^{-9}$        & $<9.2 \times 10^{-10}$        & $(6.5\pm2.4) \times 10^{-10}$\\
B & $(1.3\pm0.4) \times 10^{-6}$ & $(2.9\pm1.1) \times 10^{-9}$ & $(5.6\pm2.4) \times 10^{-9}$ & $<6.1 \times 10^{-10}$        & $(6.9\pm3.6) \times 10^{-10}$\\
C & $(1.5\pm0.5) \times 10^{-6}$ & $(3.0\pm1.3) \times 10^{-9}$ & $(6.6\pm2.7) \times 10^{-9}$ & $(1.1\pm0.3) \times 10^{-9}$  & $(6.3\pm2.3) \times 10^{-10}$\\
D & $(1.5\pm0.5) \times 10^{-6}$ & $(1.7\pm1.0) \times 10^{-9}$ & $(6.4\pm3.6) \times 10^{-9}$ & $<9.2 \times 10^{-10}$        & $(7.4\pm3.6) \times 10^{-10}$\\
E & $(1.5\pm0.6) \times 10^{-6}$ & $< 1.2 \times 10^{-9}$       & $<5.5 \times 10^{-9}$        & $<1.5 \times 10^{-9}$         & $<6.9 \times 10^{-10}$\\
\enddata
\tablecomments{Errors of the column densities are estimated by taking into account the r.~m.~s. noise and calibration uncertainty (20~\%).}
\label{tab_fra}
\end{deluxetable}

\begin{deluxetable}{lccccccc}
\tabletypesize{\scriptsize}
\tablecaption{Correlation matrix}
\tablewidth{0pt}
\tablehead{
\colhead{Map name} & \colhead{SFR} & \colhead{$^{13}$CO} & \colhead{C$^{18}$O} & \colhead{CS} & \colhead{CN} & \colhead{CH$_3$OH} & \colhead{HNCO} 
}
\startdata
SFR        & 1.0000 &        &        &        &        &        &        \\
$^{13}$CO  & 0.4912 & 1.0000 &        &        &        &        &        \\
C$^{18}$O  & 0.4301 & 0.9204 & 1.0000 &        &        &        &        \\
CS         & 0.4870 & 0.7901 & 0.7607 & 1.0000 &        &        &        \\
CN         & 0.5132 & 0.7556 & 0.6990 & 0.6889 & 1.0000 &        &        \\
CH$^{3}$OH & 0.0435 & 0.6093 & 0.5976 & 0.5817 & 0.5551 & 1.0000 &        \\
HNCO       & 0.1202 & 0.3472 & 0.3919 & 0.3723 & 0.3855 & 0.3545 & 1.0000 \\
\enddata
\label{tab_corr}
\end{deluxetable}

\begin{deluxetable}{lrrrrrrr}
\tabletypesize{\scriptsize}
\tablecaption{Eigenvectors derived in the principal component analysis}
\tablewidth{0pt}
\tablehead{
\colhead{Map name} & \colhead{PC1} & \colhead{PC2} & \colhead{PC3} & \colhead{PC4} & \colhead{PC5} & \colhead{PC6} & \colhead{PC7} 
}
\startdata
SFR                         &  0.2674 & -0.7293 & -0.2900 &  0.2289 & -0.2694 &  0.4313 & -0.0402 \\
$^{13}$CO                   &  0.4516 & -0.0617 &  0.1707 & -0.3364 &  0.2379 &  0.1283 &  0.7596 \\
C$^{18}$O                   &  0.4407 &  0.0074 &  0.1318 & -0.5163 &  0.2661 &  0.2086 & -0.6383 \\
CS                          &  0.4264 & -0.0549 &  0.0753 & -0.1477 & -0.6293 & -0.6247 & -0.0377 \\
CN                          &  0.4159 & -0.0890 & -0.0269 &  0.5860 &  0.5525 & -0.4010 & -0.0942 \\
CH$^{3}$OH                  &  0.3403 &  0.4947 &  0.3786 &  0.4463 & -0.3127 &  0.4455 & -0.0269 \\
HNCO                        &  0.2444 &  0.4568 & -0.8483 & -0.0587 & -0.0316 &  0.0668 &  0.0550 \\
Eigenvalue percentage (\%)  & 61.1    & 14.7    & 10.4    &  5.2    &  4.2    &  3.4    &  1.0    \\
\enddata
\tablecomments{Eigenvalue percentages indicate the fractions of the total variance treated by each PC component.}
\label{tab_eigv}
\end{deluxetable}


\begin{thebibliography}{}
\bibitem[Aalto et al.(1999)]{aalto1999} Aalto, S., H{\"u}ttemeister, S., Scoville, N.~Z., \& Thaddeus, P.\ 1999, \apj, 522, 165
\bibitem[Aikawa et al.(2008)]{Aikawa2008} Aikawa, Y., Wakelam, V., Garrod, R.~T., \& Herbst, E.\ 2008, \apj, 674, 984
\bibitem[Aladro et al.(2011)]{aladro2011} Aladro, R., Mart{\'{\i}}n, S., Mart{\'{\i}}n-Pintado, J., et al.\ 2011, \aap, 535, A84 
\bibitem[Aladro et al.(2013)]{aladro2013} Aladro, R., Viti, S., Bayet, E., et al.\ 2013, \aap, 549, A39 
\bibitem[Aladro et al.(2015)]{Aladro2015} Aladro, R., Mart{\'{\i}}n, S., Riquelme, D., et al.\ 2015, \aap, 579, A101 
\bibitem[Armstrong \& Barrett(1985)]{Armstrong1985} Armstrong, J.~T., \& Barrett, A.~H.\ 1985, \apjs, 57, 535
\bibitem[Bachiller \& P{\'e}rez Guti{\'e}rrez(1997)]{Bachiller1997} Bachiller, R., \& P{\'e}rez Guti{\'e}rrez, M.\ 1997, \apjl, 487, L93 
\bibitem[Bisschop et al.(2007)]{Bisschop2007} Bisschop, S.~E., J{\o}rgensen, J.~K., van Dishoeck, E.~F., \& de Wachter, E.~B.~M.\ 2007, \aap, 465, 913 
\bibitem[Calzetti et al.(2007)]{Calzetti2007} Calzetti, D., Kennicutt, R.~C., Engelbracht, C.~W., et al.\ 2007, \apj, 666, 870 
\bibitem[Colombo et al.(2014)]{Colombo2014} Colombo, D., Meidt, S.~E., Schinnerer, E., et al.\ 2014, \apj, 784, 4
\bibitem[Costagliola et al.(2011)]{costagliola2011} Costagliola, F., Aalto, S., Rodriguez, M.~I., et al.\ 2011, \aap, 528, A30 
\bibitem[Cummins et al.(1986)]{Cummins1986} Cummins, S.~E., Linke, R.~A., \& Thaddeus, P.\ 1986, \apjs, 60, 819
\bibitem[Feldmeier et al.(1997)]{feldmeier1997} Feldmeier, J.~J., Ciardullo, R., \& Jacoby, G.~H.\ 1997, \apj, 479, 231 
\bibitem[Fuente et al.(2006)]{Fuente2006} Fuente, A., Garc{\'{\i}}a-Burillo, S., Gerin, M., et al.\ 2006, \apjl, 641, L105 
\bibitem[Fujimoto(1968)]{Fujimoto1968} Fujimoto, M.\ 1968, Non-stable Phenomena in Galaxies, 29, 453 
\bibitem[Garrod et al.(2007)]{Garrod2007} Garrod, R.~T., Wakelam, V., \& Herbst, E.\ 2007, \aap, 467, 1103 
\bibitem[Ginard et al.(2015)]{Ginard2015} Ginard, D., Fuente, A., Garc{\'{\i}}a-Burillo, S., et al.\ 2015, \aap, 578, A49 
\bibitem[Helfer et al.(2003)]{helfer2003} Helfer, T.~T., Thornley, M.~D., Regan, M.~W., et al.\ 2003, \apjs, 145, 259 
\bibitem[Hughes et al.(2013)]{Hughes2013} Hughes, A., Meidt, S.~E., Schinnerer, E., et al.\ 2013, \apj, 779, 44 
\bibitem[Izumi et al.(2013)]{Izumi2013} Izumi, T., Kohno, K., Mart{\'{\i}}n, S., et al.\ 2013, \pasj, 65, 100 
\bibitem[Jackson et al.(1984)]{Jackson1984} Jackson, J.~M., Armstrong, J.~T., \& Barrett, A.~H.\ 1984, \apj, 280, 608 
\bibitem[J{\o}rgensen et al.(2005)]{Jorgensen2005} J{\o}rgensen, J.~K., Sch{\"o}ier, F.~L., \& van Dishoeck, E.~F.\ 2005, \aap, 437, 501 \bibitem[Kennicutt et al.(2003)]{Kennicutt2003} Kennicutt, R.~C., Jr., Armus, L., Bendo, G., et al.\ 2003, \pasp, 115, 928 
\bibitem[Koda et al.(2009)]{koda09} Koda, J., Scoville, N., Sawada, T., et al.\ 2009, \apjl, 700, L132 
\bibitem[Kuno et al.(2007)]{Kuno2007} Kuno, N., Sato, N., Nakanishi, H., et al.\ 2007, \pasj, 59, 117
\bibitem[Marcelino et al.(2009)]{Marcelino2009} Marcelino, N., Cernicharo, J., Tercero, B., \& Roueff, E.\ 2009, \apjl, 690, L27
\bibitem[Marcelino et al.(2010)]{Marcelino2010} Marcelino, N., Br{\"u}nken, S., Cernicharo, J., et al.\ 2010, \aap, 516, A105 
\bibitem[Mart{\'{\i}}n et al.(2006)]{martin2006} Mart{\'{\i}}n, S., Mauersberger, R., Mart{\'{\i}}n-Pintado, J., Henkel, C., \& Garc{\'{\i}}a-Burillo, S.\ 2006, \apjs, 164, 450
\bibitem[Mart{\'{\i}}n et al.(2008)]{Martin2008} Mart{\'{\i}}n, S., Requena-Torres, M.~A., Mart{\'{\i}}n-Pintado, J., \& Mauersberger, R.\ 2008, \apj, 678, 245
\bibitem[Mart{\'{\i}}n et al.(2015)]{Martin2015} Mart{\'{\i}}n, S., Kohno, K., Izumi, T., et al.\ 2015, \aap, 573, A116
\bibitem[McElroy et al.(2013)]{McElroy2013} McElroy, D., Walsh, C., Markwick, A.~J., et al.\ 2013, \aap, 550, A36 
\bibitem[Meier \& Turner(2005)]{Meier2005} Meier, D.~S., \& Turner, J.~L.\ 2005, \apj, 618, 259
\bibitem[Meier \& Turner(2012)]{Meier2012} Meier, D.~S., \& Turner, J.~L.\ 2012, \apj, 755, 104 
\bibitem[Meier et al.(2015)]{Meier2015} Meier, D.~S., Walter, F., Bolatto, A.~D., et al.\ 2015, \apj, 801, 63 
\bibitem[Miyamoto et al.(2014)]{Miyamoto2014} Miyamoto, Y., Nakai, N., \& Kuno, N.\ 2014, \pasj, 66, 36 
\bibitem[Nakai et al.(1994)]{nakai1994} Nakai, N., Kuno, N., Handa, T., \& Sofue, Y.\ 1994, \pasj, 46, 527 
\bibitem[Nakajima et al.(2011)]{nakajima2011} Nakajima, T., Takano, S., Kohno, K., \& Inoue, H.\ 2011, \apjl, 728, L38 
\bibitem[Nakajima et al.(2015)]{Nakajima2015} Nakajima, T., Takano, S., Kohno, K., et al.\ 2015, \pasj, 67, 8 
\bibitem[Pety et al.(2005)]{Pety2005} Pety, J., Teyssier, D., Foss{\'e}, D., et al.\ 2005, \aap, 435, 885
\bibitem[Roberge et al.(1991)]{Roberge1991} Roberge, W.~G., Jones, D., Lepp, S., \& Dalgarno, A.\ 1991, \apjs, 77, 287 
\bibitem[Roberts(1969)]{Roberts1969} Roberts, W.~W.\ 1969, \apj, 158, 123 
\bibitem[Rodr{\'{\i}}guez-Fern{\'a}ndez et al.(2010)]{Rodoriguez2010} Rodr{\'{\i}}guez-Fern{\'a}ndez, N.~J., Tafalla, M., Gueth, F., \& Bachiller, R.\ 2010, \aap, 516, A98
\bibitem[Sakamoto et al.(2014)]{Sakamoto2014} Sakamoto, K., Aalto, S., Combes, F., Evans, A., \& Peck, A.\ 2014, \apj, 797, 90 
\bibitem[Schinnerer et al.(2010)]{Schinnerer2010} Schinnerer, E., Wei{\ss}, A., Aalto, S., \& Scoville, N.~Z.\ 2010, \apj, 719, 1588 
\bibitem[Schinnerer et al.(2013)]{Schinnerer2013} Schinnerer, E., Meidt, S.~E., Pety, J., et al.\ 2013, \apj, 779, 42 
\bibitem[Schuster et al.(2007)]{schuster2007} Schuster, K.~F., Kramer, C., Hitschfeld, M., Garc{\'{\i}}a-Burillo, S., \& Mookerjea, B.\ 2007, \aap, 461, 143
\bibitem[Shu et al.(1972)]{Shu1972} Shu, F.~H., Milione, V., Gebel, W., et al.\ 1972, \apj, 173, 557
\bibitem[Snell et al.(2011)]{Snell2011} Snell, R.~L., Narayanan, G., Yun, M.~S., et al.\ 2011, \aj, 141, 38 
\bibitem[Soma et al.(2015)]{Soma2015} Soma, T., Sakai, N., Watanabe, Y., \& Yamamoto, S.\ 2015, \apj, 802, 74
\bibitem[Suzuki et al.(1992)]{Suzuki1992} Suzuki, H., Yamamoto, S., Ohishi, M., et al.\ 1992, \apj, 392, 551 
\bibitem[Takano et al.(2014)]{Takano2014} Takano, S., Nakajima, T., Kohno, K., et al.\ 2014, \pasj, 66, 75
\bibitem[Tercero et al.(2010)]{Tercero2010} Tercero, B., Cernicharo, J., Pardo, J.~R., \& Goicoechea, J.~R.\ 2010, \aap, 517, A96 
\bibitem[Ungerechts et al.(1997)]{Ungerechts1997} Ungerechts, H., Bergin, E.~A., Goldsmith, P.~F., et al.\ 1997, \apj, 482, 245 
\bibitem[Usero et al.(2006)]{Usero2006} Usero, A., Garc{\'{\i}}a-Burillo, S., Mart{\'{\i}}n-Pintado, J., Fuente, A., \& Neri, R.\ 2006, \aap, 448, 457 
\bibitem[Vink{\'o} et al.(2012)]{vink12} Vink{\'o}, J., Tak{\'a}ts, K., Szalai, T., et al.\ 2012, \aap, 540, A93 
\bibitem[Watanabe et al.(2014)]{Watanabe2014} Watanabe, Y., Sakai, N., Sorai, K., \& Yamamoto, S.\ 2014, \apj, 788, 4 
\bibitem[Watanabe et al.(2015)]{Watanabe2015} Watanabe, Y., Sakai, N., L{\'o}pez-Sepulcre, A., et al.\ 2015, \apj, 809, 162 
\bibitem[Zinchenko et al.(2000)]{Zinchenko2000} Zinchenko, I., Henkel, C., \& Mao, R.~Q.\ 2000, \aap, 361, 1079 
\end{thebibliography}
\end{document}